\documentclass[%
reprint,
groupedaddress,
nofootinbib,
 amsmath,amssymb,
 aps,
prb,
floatfix,
]{revtex4-2}

\usepackage{graphicx}
\usepackage{dcolumn}
\usepackage{bm}
\usepackage{braket}
\usepackage{physics}
\usepackage{comment}
\usepackage{color}
\usepackage[version=3]{mhchem}
\usepackage{ulem}

\usepackage[whole]{bxcjkjatype} 

\begin{document}

\preprint{APS/123-QED}

\title{Nonlinear spin current of photoexcited magnons in collinear antiferromagnets}

\author{Kosuke Fujiwara}
\affiliation{Department of Applied Physics, The University of Tokyo, Hongo, Tokyo, 113-8656, Japan}

\author{Sota~Kitamura}
\affiliation{Department of Applied Physics, The University of Tokyo, Hongo, Tokyo, 113-8656, Japan}

\author{Takahiro~Morimoto}
\affiliation{Department of Applied Physics, The University of Tokyo, Hongo, Tokyo, 113-8656, Japan}

\date{\today}

\begin{abstract}
We study the nonlinear magnon spin current induced by an ac electric field under light irradiation in collinear antiferromagnets with broken inversion symmetry. 
For linearly polarized light, we find that a dc spin current appears through ``the magnon spin shift current" mechanism, which is driven by a spin polarization generation in the two magnon creation process and has a close relationship to the geometry of magnon bands through Berry connection.  
For circularly polarized light,
a dc spin current appears through ``the spin injection current" mechanism, which is proportional to the relaxation time of magnons and can be large when the magnon lifetime is long.
We demonstrate the generation of the magnon spin shift and injection currents, based on a few toy models and a realistic model for a multiferroic material \ce{\textit{M}_2Mo_3O_8}.
\end{abstract}

\maketitle


\section{\label{sec:level1}Introduction \protect}
Spin transport plays a central role in reseaches of spintronics~\cite{Zutic2004Spintronics:Applications,Bader2010Spintronics}. 
In particular, the magnon transport is attracting a keen attention since magnons have a long lifetime and are able to transfer energy and spin angular momentum without Joule heating. 
Utilizing these advantages of magnons led to a research field of magnon spintronics~\cite{Chumak2015MagnonSpintronics}. 
Typical methods to create magnons in spin systems include spin pumping with an application of a microwave~\cite{Kajiwara2010TransmissionInsulator,Heinrich2011SpinInterfaces} 
and thermal responses by the application of a temperature gradient~\cite{Xiao2010TheoryEffect,Uchida2010SpinInsulator,Rezende2014MagnonEffect,Katsura2010TheoryMagnets,Onose2010ObservationEffect,Ideue2012EffectInsulators,Owerre2017TopologicalAntiferromagnets,Owerre2017TopologicalLattice,Shiomi2017ExperimentalMnPS3,Doki2018SpinAntiferromagnet,Zhang2018Spin-NernstInsulator,Kim2019MagnonAntiferromagnet,Kawano2019ThermalAntiferromagnet,Mook2019ThermalAntiferromagnets,Park2020ThermalInteraction,Fujiwara2022ThermalSystems}.
Among the thermal responses of magnons, 
the thermal Hall responses are closely related to the geometry of the magnon band.
For example, the Berry curvature of the magnon bands induce the thermal Hall effect and the spin Nernst effect ~\cite{Matsumoto2011TheoreticalFerromagnets,Matsumoto2011RotationalEffect,Matsumoto2014ThermalInteraction,Cheng2016SpinAntiferromagnets,Zyuzin2016MagnonAntiferromagnets}, and the Berry curvature dipole of the magnon bands leads to the nonlinear spin Nernst effect~\cite{Kondo2022NonlinearCurrent}.

Besides the spin pumping and thermal responses, it has been proposed that photoirradiation generates magnon spin current through a nonlinear response~\cite{Proskurin2018ExcitationInsulators,Ishizuka2019TheoryInsulators,Ishizuka2022LargeTrihalides}. 
Such nonlinear magnon spin current is analogous to a nonlinear current response of electrons to an external electric field. 
Electron systems with broken inversion symmetry exhibit photovoltaic effects, such as shift current~\cite{vonBaltz1981TheoryCrystals,Sipe2000Second-orderSemiconductors,Young2012FirstFerroelectrics,Morimoto2016TopologicalSolids,Ogawa2017ShiftSbSI} and injection current~\cite{Sipe2000Second-orderSemiconductors,deJuan2017QuantizedSemimetals,Orenstein2021TopologyResponses}. 
In particular, shift current is governed by a geometric quantity called the shift vector which quantifies the shift of the Bloch wave packet in optical transition.
Similarly, in spin systems, the application of gigahertz (GHz) or terahertz (THz) laser light can create magnons and leads to magnon spin currents. 
The application of circularly polarization light generates magnon excitations through injection of angular momentum to spin systems~\cite{Proskurin2018ExcitationInsulators}. Furthermore, the linearly polarized light is predicted to generate the magnon spin current even without angular-momentum transfer~\cite{Ishizuka2019TheoryInsulators,Ishizuka2022LargeTrihalides}.
Generation of those magnon spin currents relies on a coupling of spins to the magnetic field component of light,
because magnons are charge neutral and their coupling to the electric field is not considered usually. 
Since the magnetic field of light is small, large spin current responses through the above mechanisms require high intensity of light.

Spin systems with a broken inversion symmetry generally support electrical polarization, exhibiting a multiferroic nature~\cite{Eerenstein2006MultiferroicMaterials,Tokura2014MultiferroicsOrigin}. 
As a consequence, an electric field can directly couple with spins and generate magnetic excitations. 
For example, nonlinear responses to the electric field were studied using the spinon description for 1D systems, which includes dc spin current generation~\cite{Ishizuka2019RectificationWaves,Ikeda2019GenerationAntiferromagnet} and high harmonic generations~\cite{Ikeda2019High-harmonicMagnetization, Kanega2021LinearLiquids}. 
In higher dimensions, the low-energy excitations of the ordered magnets are usually magnons, and the magnons in multiferroic materials accompany electric polarization, known as electromagnons~\cite{Pimenov2006PossibleManganites,ValdesAguilar2009OriginRMnO3,Takahashi2012MagnetoelectricHelimagnet}.
In particular, electromagnons can be optically excited through their coupling to the electric field, leading to optical magnetoelectric effects, such as directional dichroism~\cite{Takahashi2012MagnetoelectricHelimagnet,Kezsmarki2011EnhancedCompound,Bordacs2012ChiralityExcitations}, and can be potentially applied for electric field control of magnetic orders~\cite{Mochizuki2010TheoreticallyMultiferroics,Kubacka2014Large-AmplitudeElectromagnon}.
Recently, the application of circularly polarized light was predicted to generate spin current via a two magnon Raman process with the coupling to the electric field ~\cite{Bostrom2021All-opticalEffect}.
Also, it was predicted that optical excitation of electromagnons supports the electric current generation through the shift current mechanism~\cite{Morimoto2021ElectricMagnets}.
Yet, the magnon transport induced by the electric field has not been fully explored.
In particular, the relationship between the nontrivial geometry of magnon bands and nonlinear magnon current responses is still unclear.

\begin{figure}[tb]
\centering
\includegraphics[width=\linewidth]{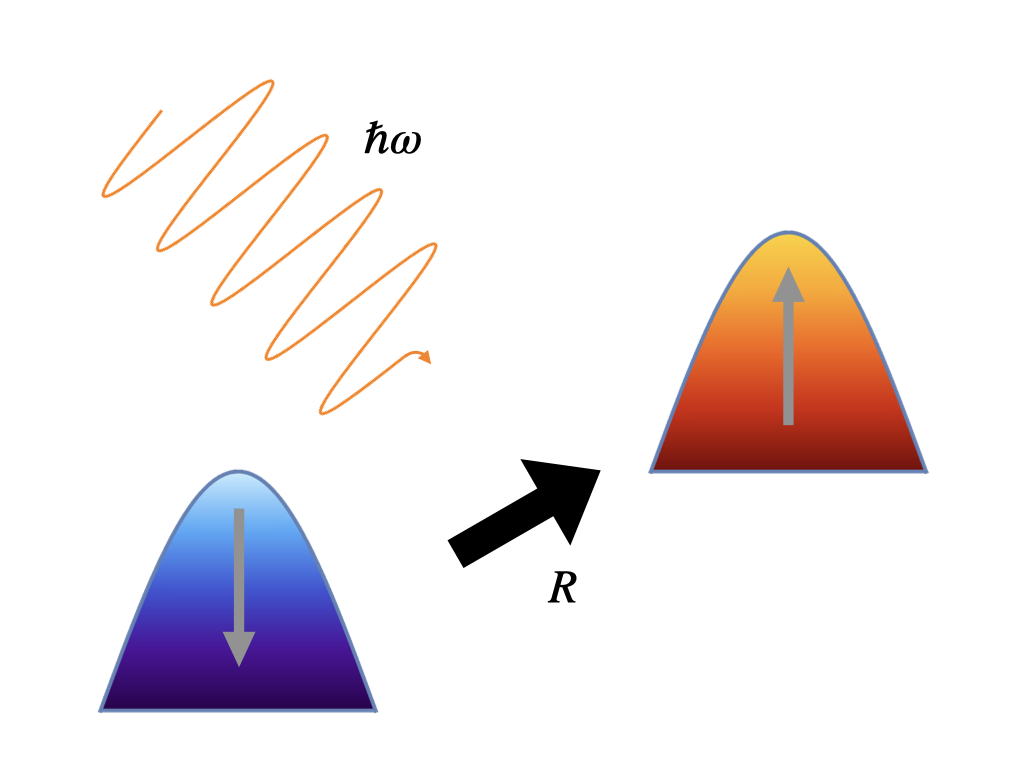}
\caption{The schematic picture of the magnon spin shift current. The electric field creates a pair of magnons with up (red) and down spins (blue). The shift vector $R$ measures the positional displacement of magnons with up and down spins.}
\label{fig:shiftvector}
\end{figure}

In this paper, we study the magnon spin current induced by the electric field in collinear antiferromagnets with broken inversion symmetry. 
Here, we focus on the dc spin current responses, and we derive the formula for the magnon spin current induced by the linearly and circularly polarized lights, using the Holstein-Primakoff transformation and the resulting magnon Hamiltonian.
For linearly polarized light, we find that a dc spin current appears through ``the magnon spin shift current" mechanism.
The magnon spin shift current can be described by a geometric quantity called the shift vector of the magnon band which represents the positional shift of magnons. 
In addition to the shift of magnons of the same spin in the usual interband transitions,
the shift vector of magnons also incorporates the positional shift between the up-spin magnon and the down-spin magnon associated with the 2-magnon excitation,  which is schematically illustrated in Fig.~\ref{fig:shiftvector}
and becomes dominant in the low-temperature limit.
For circularly polarized light,
a dc spin current appears through ``the spin injection current" mechanism.
The spin injection current is proportional to the relaxation time which is expected to exhibit a large spin current response when the magnon lifetime is long.
We demonstrate the magnon spin shift current and injection current numerically based on a few toy models and realistic models for multiferroics.

The rest of the paper is organized as follows. In Sec.~\ref{Sec:formalism}, we study the nonlinear magnon spin conductivity and in particular obtain nonlinear conductivities for the magnon spin shift and injection currents. 
We present numerical results in Secs.~\ref{sec:2d} and \ref{sec:m2mo308}. In Sec.~\ref{sec:2d}, we apply our theory to a Heisenberg model with alternating coupling constants. In Sec.~\ref{sec:m2mo308}, we study the magnon spin shift current in a more realistic model for the multiferroic material \ce{\textit{M}_2Mo_3O_8}. 
In Sec.~\ref{sec:discussion}, we give a brief discussion.

\section{Formalism of nonlinear magnon response}
\label{Sec:formalism}
In this section, we derive the formulas of the dc magnon spin current induced by an external ac electric field. 
We first summarize notations and then derive the nonlinear magnon spin current conductivity for collinear magnets from a standard perturbation theory. 
Under the effective time-reversal symmetry (TRS), we obtain a concise expression for the magnon spin shift current conductivity which reveals the relationship between the magnon spin shift current and the shift vector of magnon bands that reflects a nontrivial geometry of the magnon bands. 
In addition, we derive the magnon spin injection current conductivity induced by the circularly polarized light.

\subsection{Magnon Hamiltonian and diagonalization}

We consider collinear antiferromagnets described by the spin Hamiltonian
\begin{equation}
    H_S=\sum_{i,j}J_{ij}\vb*{S}_i\cdot\vb*{S}_j+\sum_i\Delta_i (S^z_i)^2,
\end{equation}
where $\vb*{S}_i$ is a spin operator at $i$th site. We note that our theory is also applicable to more general Hamiltonian, such as the XXZ model. Hereafter, we take the z-axis parallel to the spin direction. The low-energy excitations of the ordered magnets are usually magnons, thus we consider magnons via the Holstein Primakoff transformation~\cite{Holstein1940FieldFerromagnet}
\begin{equation} \label{HPtrans}
\begin{cases} 
    S^{+}_i\simeq\hbar\sqrt{2S}a_i,
    S^{-}_i\simeq\hbar\sqrt{2S}a^\dagger_i,
    S^{z}_i=\hbar(S-a^\dagger_i a_i) \\
    ~~~~~~~~~~~~~~~~~~~~~~~~~~~~~~~~~~~~~~~~~~~~\text{for $\langle S^z_i\rangle =S$}  \\
    S^{+}_i\simeq\hbar\sqrt{2S}b^\dagger_i,
    S^{-}_i\simeq\hbar\sqrt{2S}b_i,
    S^{z}_i=\hbar(-S+b^\dagger_i b_i) \\
    ~~~~~~~~~~~~~~~~~~~~~~~~~~~~~~~~~~~~~~~~~~~~\text{for $\langle S^z_i\rangle =-S$}
\end{cases},
\end{equation}
where $S$ is a spin of $\vb*{S}_i$, and $a^\dagger_i$ and $b^\dagger_i$ are creation operators of the magnon of the $i$th site. Hereafter, we set $\hbar=1$. By using the Holstein Primakoff transformation, we obtain the magnon Hamiltonian
\begin{equation}
    \hat{H}=\sum_{\vb*{R},\vb*{R^\prime}}\Psi^\dagger_{\vb*{R}}H_{\vb*{R},\vb*{R^\prime}}\Psi_{\vb*{R^\prime}}.
\end{equation}
Here $\Psi^\dagger_{\vb*{R}}=(a^\dagger_{\vb*{R}_1}\cdots a^\dagger_{\vb*{R}_N},b_{\vb*{R}_1^\prime}\cdots b_{\vb*{R}_M^\prime})$ and $\vb*{R}$ is a position of the unit cell. $a^\dagger_{\vb*{R}_i}$ $b^\dagger_{\vb*{R}_i}$is a creation operator of the magnon of the $i$th site of the unit cell at $\vb*{R}$, and $\vb*{R}_i=\vb*{R}+\vb*{r}_i$ denotes the position of the $i$th site in the unitcell at $\vb*{R}$. The magnon Hamiltonian in the momentum space is
\begin{equation}
    \hat{H}=\sum_{\vb*{k}}\Psi^\dagger_{\vb*{k}}H_{\vb*{k}}\Psi_{\vb*{k}},
\end{equation}
where $\Psi^\dagger_{\vb*{k}}=(a^\dagger_{1\vb*{k}}\cdots a^\dagger_{N\vb*{k}},b_{1-\vb*{k}}\cdots b_{M-\vb*{k}})$. Here, $a^\dagger_i(\vb*{k})=\frac{1}{\sqrt{N}}\sum_{\vb*{R}}a^\dagger_{\vb*{R}_i}e^{i\vb*{k}\cdot(\vb*{R}+\vb*{r}_i)}$ and $b^\dagger_i(\vb*{k})=\frac{1}{\sqrt{N}}\sum_{\vb*{R}}b^\dagger_{\vb*{R}_i}e^{i\vb*{k}\cdot(\vb*{R}_i+\vb*{r_i})}$, and $N$ is the total number of unit cells.

We can diagonalize the Hamiltonian by the paraunitary matrix $V_{\vb*{k}}$ which satisfies $V^\dagger_{\vb*{k}} BV_{\vb*{k}}=B$. Here, $B$ is a diagonal matrix
\begin{equation}
    B=\textrm{diag}(\eta_a),
\end{equation}
where $\eta_a=1$ if $(\Psi^\dagger_{\vb*{k}})_a$ is a creation operator and $\eta_a=-1$ if $(\Psi^\dagger_{\vb*{k}})_a$ is a annihilation operator. Thus $B$ satisfies $(B)_{ab}=[(\Psi_{\vb*{k}})_a,(\Psi^\dagger_{\vb*{k}})_b]$ and we obtain
\begin{align}
    \hat{H}&=\sum_{\vb*{k}}\Psi^\dagger_{\vb*{k}}V^{\dagger -1}_{\vb*{k}}E_{\vb*{k}}V^{-1}_{\vb*{k}}\Psi_{\vb*{k}}\notag\\
    &=\sum_{\vb*{k}}\Phi^\dagger_{\vb*{k}}E_{\vb*{k}}\Phi_{\vb*{k}}.
\end{align}
Here, $E_{\vb*{k}}$ is diagonal matrix
\begin{equation}
E_{\vb*{k}}=V^{\dagger}_{\vb*{k}}H_{\vb*{k}}V_{\vb*{k}}
\end{equation}
and $\Phi_{\vb*{k}}$ is a transformed operator
\begin{equation}
\Phi_{\vb*{k}}=V^{-1}_{\vb*{k}}\Psi_{\vb*{k}},
\end{equation}
and $\Phi_{\vb*{k}}$ satisfies a commutation relation $[(\Phi_{\vb*{k}})_a,(\Phi^\dagger_{\vb*{k}})_b]=(B)_{ab}$. 
Matrix elements of $E_{\vb*{k}}$ are positive when the ground state is stable, and matrix elements of $E_{\vb*{k}}$ have physical meaning as the excitation energies of magnons.

Here, we consider the distribution function
\begin{equation}
    \rho_{\vb*{k}a}=\langle \Phi^\dagger_{\vb*{k}a}\Phi_{\vb*{k}a}\rangle, 
\end{equation}
where $\langle\mathcal{O}\rangle$ is the expectation value of $\mathcal{O}$ in the equilibrium state. Since $\Phi_{\vb*{k}}$ is the basis of the diagonalized form of the Hamiltonian, we obtain
\begin{equation}
\begin{cases}
    \rho_{\vb*{k}a}=1/(\exp{\beta (E_{\vb*{k}a})_{aa}}-1) \\
    ~~~~~~~~~~~~~~~~~~~~~~~~~~~~~~~~~~~~~\textrm{for}~[(\Phi_{\vb*{k}})_a,(\Phi^\dagger_{\vb*{k}})_a]=1 \\
    \rho_{\vb*{k}a}=-1/(\exp{-\beta (E_{\vb*{k}})_{aa}}-1)\\
    ~~~~~~~~~~~~~~~~~~~~~~~~~~~~~~~~~~~~~\textrm{for}~[(\Phi_{\vb*{k}})_a,(\Phi^\dagger_{\vb*{k}})_a]=-1\\
\end{cases}
\end{equation}
To simplify $\rho_{\vb*{k}a}$, we introduce $\varepsilon_{\vb*{k}}$
\begin{equation}
\varepsilon_{\vb*{k}}=BE_{\vb*{k}}=BV^{\dagger}_{\vb*{k}}H_{\vb*{k}}V_{\vb*{k}}=V^{-1}_{\vb*{k}}BH_{\vb*{k}}V_{\vb*{k}},\label{varepsilon}
\end{equation}
and we can write $\rho_{\vb*{k}a}=B_{aa}/(\exp{\beta (\varepsilon_{\vb*{k}a})}-1)$

Now, we consider the general operator
\begin{align}
    \hat{\mathcal{O}}&=\sum_{\vb*{k}}\Psi^\dagger_{\vb*{k}}\mathcal{O}_{\vb*{k}}\Psi_{\vb*{k}}\notag\\
    &=\sum_{\vb*{k}}\Phi^\dagger_{\vb*{k}}B\tilde{\mathcal{O}}_{\vb*{k}}\Phi_{\vb*{k}},
\end{align}
where we define
\begin{align}
    \tilde{\mathcal{O}}_{\vb*{k}} \equiv V^{-1}_{\vb*{k}}B\mathcal{O}_{\vb*{k}}V_{\vb*{k}}. 
    \label{eq: def O tilde}
\end{align}
Here, we note that $\tilde{O}_{\vb*{k}}$ is generally non-Hermitian matrix. However, matrix elements of $\tilde{O}_{\vb*{k}}$ satisfies
\begin{equation}
 (\tilde{O}_{\vb*{k}})_{ab}=B_{aa}B_{bb}(\tilde{O}_{\vb*{k}})_{ba}^*. 
 \label{o_tenchi}
\end{equation}

\subsection{Polarization and spin current}
Based on the above conventions, let us study the nonlinear spin current induced by an external light field. 
The electric field $\vb*{E}$ creates magnon excitations via the coupling to the electric polarization in magnetic systems.
The total Hamiltonian in the presence of the external electric field can be written as
\begin{equation}
    \hat{H}_{tot}=\hat{H}-\vb*{E}\cdot\hat{\vb*{P}},
    \label{total_ham}
\end{equation}
where $\hat{\vb*{P}}$ is the polarization operator
\begin{equation}
    \hat{\vb*{P}}=-\sum_{\vb*{k}}\Psi^\dagger_{\vb*{k}}\vb*{\Pi}_{\vb*{k}}\Psi_{\vb*{k}}.
\end{equation}
Thus the spin current $\vb*{J}$ defined via the continuity equation $-\nabla\vb*{J}=\partial_t S^z=-i[S^z,H]$ is expressed as~\cite{Zyuzin2016MagnonAntiferromagnets,Proskurin2018ExcitationInsulators}
\begin{align}
\label{spin_current}
    \hat{J}^\mu&=\sum_{\vb*{k}}\Psi^\dagger_{\vb*{k}}\frac{\partial H_{\vb*{k}}}{\partial k_\mu}\Psi_{\vb*{k}}+\sum_{\vb*{k}}\Psi^\dagger_{\vb*{k}}\frac{\partial\vb*{\Pi}_{\vb*{k}}\cdot\vb*{E}}{\partial k_\mu}\Psi_{\vb*{k}}\notag\\
    &=\hat{J}_1^\mu+\sum_\alpha \hat{J}_2^{\mu\alpha}E_\alpha.
\end{align}
Here, we decompose $\hat{J}^\mu$ into $\hat{J}_1^\mu$ and $\hat{J}^\mu_2$. $\hat{J}_1^\mu$ is a term consisting of the $k$-derivative of $\hat{H}_{\vb*{k}}$ and $\hat{J}^\mu_2$ is a term consisting of the $k$-derivative of external field $\vb*{E}\cdot\vb*{\Pi}_{\vb*{k}}$. The nonlinear magnon spin current response $J^{\mu(2)}$ can be written as
\begin{equation}
    \langle J^{\mu}(\omega)\rangle=\sigma^{\mu,\alpha\beta}(\omega,\omega_1,\omega_2)E_\alpha(\omega_1)E_\beta(\omega_2),
\end{equation}

By using the Green function formalism (for details, see Appendix~\ref{Appendix}), we obtain the nonlinear magnon spin conductivity $\sigma^{\mu,\alpha\beta}(0,\omega,-\omega)$ as 
\begin{widetext}
\begin{align}
    &\sigma^{\mu,\alpha\beta}(0,\omega,-\omega)=\notag\\
    &-\int\frac{d\vb*{k}}{(2\pi)^d}\left[\sum_{a,b}\tilde{J_2}^{\mu\alpha}_{ab}\tilde{\Pi}^\beta_{ba} \frac{f_{ab}}{\varepsilon_{ab}+\omega+i\delta}+\sum_{a,b,c}\frac{\tilde{J_1}^\mu_{ac}\tilde{\Pi}^\alpha_{cb}\tilde{\Pi}^\beta_{ba}}{\varepsilon_{ac}+2i\delta}\left( \frac{f_{ab}}{\varepsilon_{ab}+\omega+i\delta}+\frac{f_{cb}}{\varepsilon_{bc}-\omega+i\delta}\right)\right]+(\alpha,\omega\leftrightarrow\beta,-\omega).
    \label{2ndsigma}
\end{align}
\end{widetext}
Here, $\varepsilon_{ab}=\varepsilon_{\vb*{k}a}-\varepsilon_{\vb*{k}b}$ is the difference of the band dispersion and $f_{ab}=f(\varepsilon_{\vb*{k}a})-f(\varepsilon_{\vb*{k}b})$, where $f(\varepsilon_{\vb*{k}a})=(\exp{\beta\varepsilon_{\vb*{k}a}}-1)^{-1}=B_{aa}\rho_{\vb*{k}a}$. In the low temperature limit, we obtain $\rho_{\vb*{k}a}=0$ when $\varepsilon_{\vb*{k}a}$ is positive and $\rho_{\vb*{k}a}=1$ when $\varepsilon_{\vb*{k}a}$ is negative.
Here we note that we use $\varepsilon_{\vb*{k}a}$ instead of using the excitation energy $E_{\vb*{k}}=B\varepsilon_{\vb*{k}}$. Thus we have a formulation which counts states with $\varepsilon_{\vb*{k}a}<0$ by "negative counts" via $f(\varepsilon_{\vb*{k}a})=B_{aa}\rho_{\vb*{k}a}$.
The last term of (\ref{2ndsigma}) diverges as $\propto 1/\delta$ when $a=c$ and is analog of the magnon spin injection current.

\subsection{Magnon spin shift current}
\label{subsec:shift}
Now, we consider the shift current
\begin{equation}
    J_{\textrm{shift}}^{\mu}(\omega)=\sigma^{\mu,\alpha\alpha}_{\textrm{shift}}(0,\omega,-\omega)E_\alpha(\omega)E_\alpha(-\omega),
\end{equation}
under the effective TRS: $H_{\vb*{k}}=H^*_{-\vb*{k}}$ and $\Pi^\alpha_{\vb*{k}}=(\Pi^{\alpha}_{-\vb*{k}})^*$.
From the effective TRS, matrix elements satisfy
\begin{subequations}
\label{effectiveTRS}
\begin{gather}
    \tilde{J}^\mu_{1\vb*{k}}=-(\tilde{J}^\mu_{1-\vb*{k}})^*,\\
    \tilde{J}^{\mu\alpha}_{2\vb*{k}}=-(\tilde{J}^{\mu\alpha}_{2-\vb*{k}})^*,\\
    \varepsilon_{\vb*{k}a}=\varepsilon_{-\vb*{k}a}.
\end{gather}
\end{subequations}
Using Eqs.~(\ref{effectiveTRS}) and the relation $\frac{1}{x+i\delta}=\mathcal{P}\frac{1}{x}-i\pi\delta(x)$ with $\mathcal{P}$ representing the principal value, terms containing principal value are odd in $\vb*{k}$ and vanish. Thus we obtain
\begin{align}
&\sigma^{\mu,\alpha\alpha}_{\textrm{shift}}(0,\omega,-\omega)=\notag\\
&-\pi\int\frac{d\vb*{k}}{(2\pi)^d}\Biggl[\sum_{ab}\textrm{Im}[\tilde{J_2}_{ab}^{\mu\alpha}\tilde{\Pi}^\alpha_{ba}]f_{ab}\delta(\varepsilon_{ab}-\omega)\notag\\
&+\sum_{a,b,c}\frac{\textrm{Im}\left[\tilde{J_1}^\mu_{ac}\tilde{\Pi}^\alpha_{cb}\tilde{\Pi}^\alpha_{ba}\right]}{\varepsilon_{ac}+2i\delta}\left(f_{ab}\delta(\varepsilon_{ab}+\omega)+f_{cb}\delta(\varepsilon_{bc}-\omega)\right)\Biggr]\notag\\
&+(\omega\leftrightarrow-\omega).
\label{shift_tochu}
\end{align}
The last term vanishes when $a=c$ under the effective TRS, and we can remove $2i\delta$ in the denominator of the last term. By using Eq.~(\ref{o_tenchi}), Eq.~(\ref{shift_tochu}) can be written as
\begin{align}
&\sigma^{\mu,\alpha\alpha}_{\textrm{shift}}(0,\omega,-\omega)=\notag\\
&-2\pi\int\frac{d\vb*{k}}{(2\pi)^d}\Biggl[\sum_{ab}\textrm{Im}[\tilde{J_2}_{ab}^{\mu\alpha}\tilde{\Pi}^\alpha_{ba}]f_{ab}\delta(\varepsilon_{ab}-\omega)\notag\\
&+\sum_{a,b,c}\textrm{Im}\left[\tilde{J_1}^\mu_{ac}\tilde{\Pi}^\alpha_{cb}\tilde{\Pi}^\alpha_{ba}\right]\frac{f_{ab} }{\varepsilon_{ac}}(\delta(\varepsilon_{ab}+\omega)+\delta(\varepsilon_{ab}-\omega))\Biggr].
\label{shift_sigma}
\end{align}

Furthermore, we can rewrite $\tilde{J_1}^\mu$ and $\tilde{J_2}^\mu$ by using the expression for the gauge covariant derivative,
\begin{align}
\widetilde{\frac{\partial\mathcal{O}_{\vb*{k}}}{\partial k_\mu}}&\equiv
V^{-1}_{\vb*{k}}B\frac{\partial\mathcal{O}_{\vb*{k}}}{\partial k_\alpha}V_{\vb*{k}}\notag \\
&={\frac{\partial\tilde{\mathcal{O}}_{\vb*{k}}}{\partial k_\mu}}-\tilde{O}_{\vb*{k}}V^{-1}_{\vb*{k}}\frac{\partial V_{\vb*{k}}}{\partial k_\mu}-\frac{\partial V^{-1}_{\vb*{k}}}{\partial k_\mu}V_{\vb*{k}}\tilde{O}_{\vb*{k}}\notag\\
&={\frac{\partial\tilde{\mathcal{O}}_{\vb*{k}}}{\partial k_\mu}}-[i\mathcal{A}^\mu_{\vb*{k}},\tilde{O}_{\vb*{k}}],
\label{berry_con}
\end{align}
where $\mathcal{A}^\mu_{\vb*{k}}=iV^{-1}_{\vb*{k}}\frac{\partial V_{\vb*{k}}}{\partial k_\mu}=iBV^{\dagger}_{\vb*{k}}B\frac{\partial V_{\vb*{k}}}{\partial k_\mu}$ is a Berry connection.
By using Eq.~(\ref{berry_con}) for $H_{\vb*{k}}$ and $\Pi_{\vb*{k}}$, we can rewrite Eq.~(\ref{shift_sigma}) as
\begin{align}
&\sigma^{\mu,\alpha\alpha}_
{\textrm{shift}}(0,\omega,-\omega)=\notag\\
&-2\pi\int\frac{d\vb*{k}}{(2\pi)^d}\sum_k\sum_{a,b}\textrm{Im}\biggl[|\tilde{\Pi}^\alpha_{ab}|^2\Bigl(\partial_{k_\mu}\ln{\tilde{\Pi}^\alpha_{ab}}\notag\\
&-i(\mathcal{A}^\mu_{aa}-\mathcal{A}^\mu_{bb})\Bigr)-\tilde{\Pi}^\alpha_{ba}(\sum_{c\neq a}i\mathcal{A}^\mu_{ac}\tilde{\Pi}^\alpha_{cb}-\sum_{c\neq b}i\mathcal{A}^\mu_{cb}\tilde{\Pi}^\alpha_{ac})\notag\\
&+\tilde{\Pi}^\alpha_{ba}(\sum_{c\neq a}i\mathcal{A}^\mu_{ac}\tilde{\Pi}^\alpha_{cb}-\sum_{c\neq b}i\mathcal{A}^\mu_{cb}\tilde{\Pi}^\alpha_{ac})\biggr]f_{ab}\delta(\varepsilon_{ab}-\omega)\notag\\
&=-2\pi\int\frac{d\vb*{k}}{(2\pi)^d}\sum_{a,b}|\tilde{\Pi}^\alpha_{ab}|^2R^{\mu}f_{ab}\delta(\varepsilon_{ab}-\omega). \label{magnon_shift}
\end{align}
Here, $R^{\mu}=\textrm{Im}[\partial_{k_\mu}\ln{\tilde{\Pi}^\alpha_{ab}}-i(\mathcal{A}^\mu_{aa}-\mathcal{A}^\mu_{bb})]$, which can be regarded as an analog of a shift vector that appears in the expression for electronic shift current in semiconductors~\cite{Sipe2000Second-orderSemiconductors}. 
This magnon shift vector contains the difference of Berry connections of two magnon bands involved in the optical transition and is a gauge invariant quantity as a whole, 
which effectively measures the spin polarization induced by the two optically created magnons.
The shift vector of magnons incorporates the positional shift between the up-spin magnon and the down-spin magnon associated with the two-magnon excitation, in addition to the shift of magnons of the same spin in the usual interband transitions (Fig.~\ref{fig:shiftvector}).
One clear difference between the magnon shift current and electronic shift current is that the expression for the electronic shift current involves the square of the absolute value of the velocity operator, the shift vector, and the Fermi distribution function, whereas Eq.~(\ref{magnon_shift}) for the magnon shift current involves the square of the absolute value of the polarization operator, magnon shift vector, and the Bose distribution function. 
This arises from the difference in the statistical properties of magnons and electrons and their couplings to the electric field. The interaction with the electric field $\vb*{E}$ is described as $\vb*{P}\cdot\vb*{E}$ in both systems, but the polarization $\vb*{P}$ differs between the electron and magnon systems. In the electronic system, $\vb*{P}$ is proportional to the position $\vb*{r}$ in the real space. However, in the magnon system, $\vb*{P}$ is defined at each bond and not necessarily proportional to the position $\vb*{r}$.

\subsection{Magnon spin injection current under the circular polarization light}
Now, we consider the injection current by focusing on the $a=c$ term in Eq.~(\ref{2ndsigma}). For linearly polarized light, this term vanishes under the effective TRS as seen in Sec.~\ref{subsec:shift}. 
Thus we consider the circularly polarized light $\vb*{E}(\omega)=E_0(\omega)\hat{\alpha}+iE_0(\omega)\hat{\beta}$ as the external field to break the effective TRS. 
Here, $\hat{\alpha}$ ($\hat{\beta}$) is a unit vector of direction $\alpha$ ($\beta$). The injection current can be described by the conductivity $\sigma_{\textrm{inj}}^{\mu,\alpha\beta}$, which is defined by
\begin{equation}
    J_{\textrm{inj}}^\mu=\sigma_{\textrm{inj}}^{\mu,\alpha\beta}E_\alpha(\omega)E_\beta(-\omega)
    ,
\end{equation}
Since we consider the circularly polarized light, the $a=c$ term in Eq.~(\ref{2ndsigma}) is given by
\begin{align*}
    &\sigma^{\mu,\alpha\beta}_{\textrm{inj}}(0,\omega,-\omega)=\\
    &-2i\int\frac{d\vb*{k}}{(2\pi)^d}\sum_{b,a=c}\frac{\tilde{J}^\mu_{ac}\tilde{\Pi}^\alpha_{cb}\tilde{\Pi}^\beta_{ba}}{\varepsilon_{ac}+2i\delta
}\biggl(\frac{f_{ab}}{\varepsilon_{ab}+\omega+i\delta}\\
&+\frac{f_{cb}}{\varepsilon_{bc}-\omega+i\delta}\biggr)+(\alpha,\omega\leftrightarrow\beta,-\omega)\\
&=-2i\int\frac{d\vb*{k}}{(2\pi)^d}\sum_{a,b}\frac{\tilde{J}^\mu_{aa}\tilde{\Pi}^\alpha_{ab}\tilde{\Pi}^\beta_{ba}}{2i\delta
}f_{ab}\\
&\times\left(\frac{1}{\varepsilon_{ab}+\omega+i\delta}-\frac{1}{\varepsilon_{ab}+\omega-i\delta}\right)+(\alpha,\omega\leftrightarrow\beta,-\omega).
\end{align*}
With the relation $\frac{1}{x+i\delta}=\mathcal{P}\frac{1}{x}-i\pi\delta(x)$, only resonant terms containing the delta function becomes nonzero and we obtain
\begin{align}
\sigma^{\mu,\alpha\beta}_{\textrm{inj}}(0,\omega,-\omega)=&-4\pi\int\frac{d\vb*{k}}{(2\pi)^d}\sum_{a,b}\frac{\tilde{J}^\mu_{aa}\tilde{\Pi}^\alpha_{ab}\tilde{\Pi}^\beta_{ba}}{2i\delta
}\notag\\
&\times
f_{ab}\delta(\varepsilon_{ab}+\omega)+(\alpha,\omega\leftrightarrow\beta,-\omega)\notag\\
=&-2\pi\tau\int\frac{d\vb*{k}}{(2\pi)^d}\sum_{a,b}(\tilde{J}_{aa}^{\mu}-\tilde{J}_{bb}^{\mu})\notag\\
&\times\textrm{Im}[\tilde{\Pi}^\beta_{ba}\tilde{\Pi}^\alpha_{ab}]f_{ab}\delta(\varepsilon_{ab}+\omega).\label{injection_current}
\end{align}
Here, we introduce the relaxation time $\tau=1/\delta$. This expression is analogous to that for the injection current in the electronic system ~\cite{Sipe2000Second-orderSemiconductors,deJuan2017QuantizedSemimetals}. 
We note that, in the electronic case, the above expression had a  geometrical meaning in the two band limit in that the term $\textrm{Im}[\tilde{\Pi}^\beta_{ba}\tilde{\Pi}^\alpha_{ab}]$ reduces to the Berry curvature of the electronic bands ~\cite{deJuan2017QuantizedSemimetals}.
In the present case, the polarization operator $\bm{P}$ is not necessarily proportional to the position operator $\bm{r}$ and the term $\textrm{Im}[\tilde{\Pi}^\beta_{ba}\tilde{\Pi}^\alpha_{ab}]$  does not have a direct relationship to the Berry curvature of magnon bands.

\section{2D model}
\label{sec:2d}
To demonstrate the magnon spin shift current and injection current, we consider an inversion-broken 2D model on a square lattice obtained as an extension of the Rice-Mele Hubbard model~\cite{Katsura2009TheoryInsulators,Morimoto2021ElectricMagnets} into the two dimensions. Specifically, we introduce a staggered potential and staggered hopping on the square-lattice Hubbard model, which leads to the broken inversion symmetry and nonvanishing polarization.

\begin{figure}[tb]
\centering
\includegraphics[width=\linewidth]{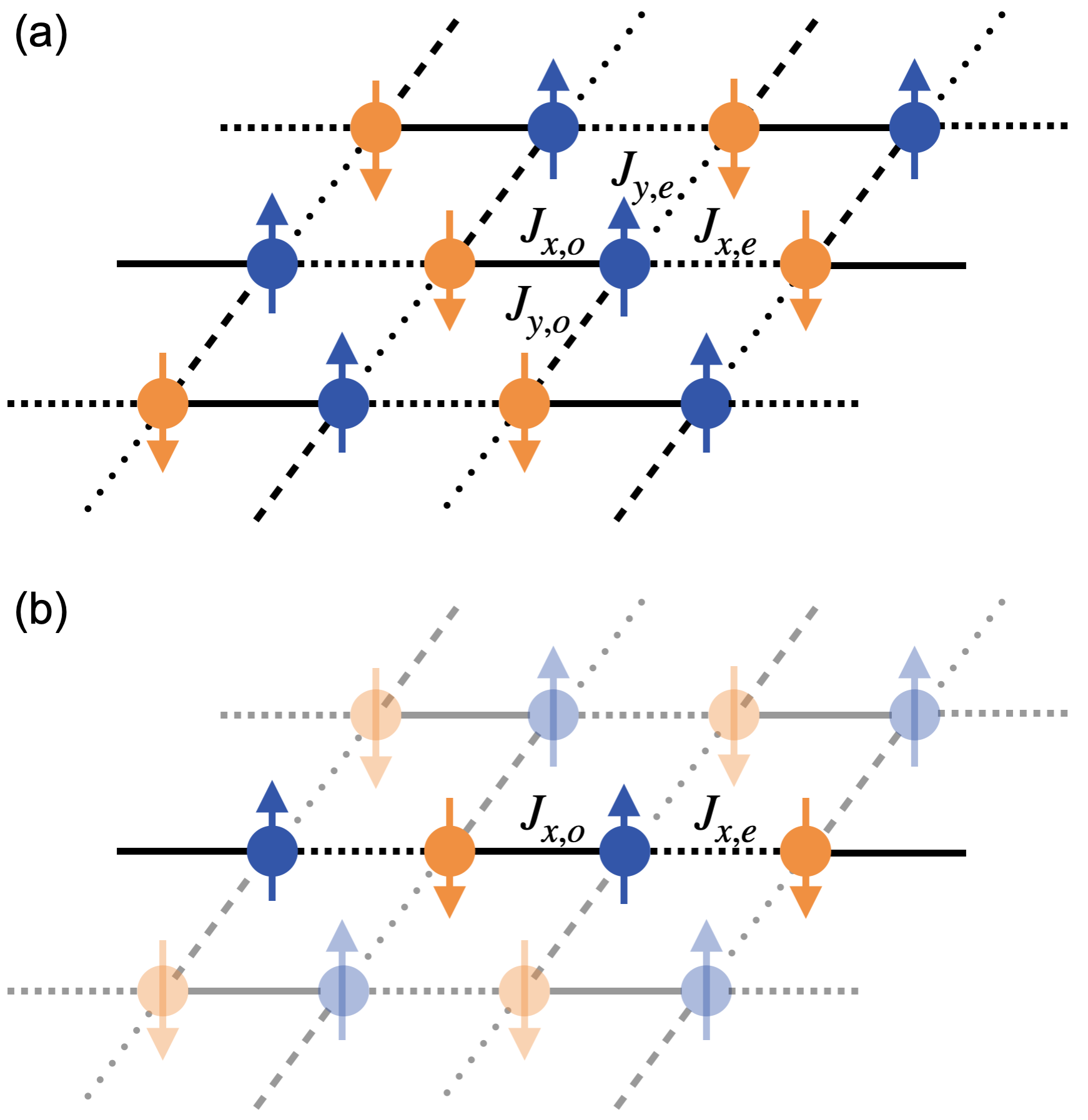}
\caption{Schematic picture of the Heisenberg model with alternating exchange interaction. (a) 2D model with 4 types of exchange interactions: $J_{x,e}$, $J_{y,e}$, $J_{x,o}$, and $J_{y,o}$. (b) Schematic picture of the 1D limit. 
when $J_{y,e}$ and $J_{y,o}$ are negligible, the 2D model reduces to a stack of Heisenberg chains with alternating couplings.}
\label{fig:2dmodel}
\end{figure}

\subsection{Definition of the model}

First, we consider the inversion broken Hubbard model on the square lattice defined by
\begin{align}
    \hat{H_0}=&\sum_{i_x,i_y,s} \left[\{t_x+(-1)^{i_x+i_y}\delta t_x\}c^\dagger_{i_x+1,i_y,s}c_{i_x,i_y,s}+h.c.\right.\notag\\
    &\left.+\{t_y+(-1)^{i_x+i_y}\delta t_y\}c^\dagger_{i_x,i_y+1,s}c_{i_x,i_y,s}+h.c.\right.\notag\\
    &\left.+(-1)^{i_x+i_y}mc^\dagger_{i_x,i_y,s}c_{i_x,i_y,s}\right]\notag\\
    &+U\sum_{i_x,i_y} n_{i_x,i_y,\uparrow}n_{i_x,i_y,\downarrow},
\end{align}
where $c_{i_x,i_y,s}$ is an annihilation operator of the electrons at $(i_x,i_y)$th site and spin $s=\uparrow,\downarrow$, and $n_{i_x,i_y,s}=c^\dagger_{i_x,i_y,s}c_{i_x,i_y,s}$ is a density operator. $t_x$ ($t_y$) is the overall hopping strength for $x$ ($y$) direction, $\delta t_x$ ($\delta t_y$) is the hopping alternation for $x$ ($y$) direction, $m$ is a staggered potential, and $U$ is a onsite Coulomb potential. For sufficiently large $U$, the ground state is in the Mott-insulating phase, and we can derive effective spin model. Figure~\ref{fig:2dmodel} shows the schematic picture of the spin model.

Now, we apply the electric field $\vb*{E}$ and consider the strong coupling expansion. Hereafter, we set electric charge $e$ as $e=1$. The electric field plays an role of a site-dependent onsite potential, and we obtain the effective spin model as shown in Fig.~\ref{fig:2dmodel}(a) as
\begin{align}
    \hat{H}_S=&\sum_{i_x,i_y}\left[J_{x,i}\vb*{S}_{i_x+1,i_y}\cdot\vb*{S}_{i_x,i_y}+J_{y,i}\vb*{S}_{i_x,i_y+1}\cdot\vb*{S}_{i_x,i_y}\right]\notag\\
    &+\sum_{i_x,i_y}\left[E_x\Pi_{x,i}\vb*{S}_{i_x+1,i_y}\cdot\vb*{S}_{i_x,i_y}\right.\notag\\
    &\left.+E_y\Pi_{y,i}\vb*{S}_{i_x,i_y+1}\cdot\vb*{S}_{i_x,i_y}\right]+\mathcal{O}(E^2),
\end{align}
where $i$ is a simplified notation of $(i_x,i_y)$ and
\begin{subequations}
\begin{align}
    J_{\alpha,i}=&2(t_\alpha+(-1)^{i_x+i_y}\delta t_\alpha)^2\left(\frac{1}{U-2m}+\frac{1}{U+2m}\right),\\
    \Pi_{\alpha,i}=&2a_\alpha(t_\alpha+(-1)^{i_x+i_y}\delta t_\alpha)^2\notag\\
    &\times\left(-\frac{(-1)^{i_x+i_y}}{(U-2m)^2}+\frac{(-1)^{i_x+i_y}}{(U+2m)^2}\right).
    \end{align}
\end{subequations}
Here, $\alpha$ denotes the direction $x$ or $y$, and $a_\alpha$ is a lattice constant. For simplification of notation, we introduce $J_\alpha=(J_{\alpha,e}+J_{\alpha,o})/2$, $\delta J_\alpha=(J_{\alpha,o}-J_{\alpha,e})/2$, $\Pi_\alpha=(\Pi_{\alpha,e}+\Pi_{\alpha,o})/2$, and $\delta \Pi_\alpha=(\Pi_{\alpha,o}-\Pi_{\alpha,e})/2$. Here $J_{\alpha,e}$ ($J_{\alpha,o}$) denotes $J_{\alpha,i}$ where $i_x+i_y$ is even (odd).

 Now, we assume $\langle S_i^z\rangle=S$ for even sites and $\langle S_i^z\rangle=-S$ for odd sites, and apply Holstein-Primakoff transformation (\ref{HPtrans}) to obtain the magnon Hamiltonian
 \begin{equation}
 \label{2d_magham}
     H_{\vb*{k}}=2S
     \begin{pmatrix}
      J_x+J_y &  \gamma_x(k_x)+\gamma_y(k_y)\\
      \gamma_x^*(k_x)+\gamma_y^*(k_y) & J_x+J_y
     \end{pmatrix}.
 \end{equation}
Here, $\gamma_{x}(k_x)=J_x\cos{(k_xa_x)}-i\delta J_x\sin{(k_xa_x)}$ and $\gamma_y(k_y)=J_y\cos{(k_ya_y)}-i\delta J_y\sin{(k_ya_y)}$. The polarization is given by
 \begin{equation}
     \Pi_{\vb*{k}}=2S
     \begin{pmatrix}
      \Pi_x+\Pi_y &  \eta_x(k_x)+\eta_y(k_y)\\
      \eta_x^*(k_x)+\eta_y^*(k_y) & \Pi_x+\Pi_y
     \end{pmatrix}.
 \end{equation}
Here, $\eta_{x}(k_x)=\Pi_x\cos{(k_xa_x)}-i\delta \Pi_x\sin{(k_xa_x)}$ and $\eta_y(k_y)=\Pi_y\cos{(k_ya_y)}-i\delta \Pi_y\sin{(k_ya_y)}$. By diagonalizing the Hamiltonian (\ref{2d_magham}) we obtain the energy dispersion $E_{\vb*{k}}=2S\sqrt{(J_x+J_y)^2-|\gamma_x(k_x)+\gamma_y(k_y)|^2}$.

\subsection{Photoinduced spin current}
Next, we study the photoinduced spin current of the shift and injection origins in this model. 
Since the spin shift current appears along the polar direction with the linearly polarized light, we consider the 1D limit of the above model as a minimum setup for spin shift current, for simplicity.
In the presence of effective TRS, the spin injection current appears with the circularly polarized light and requires the 2D nature of the model. Therefore, we treat the full 2D model to demonstrate the spin injection current.

\subsubsection{Spin shift current in the 1D limit}

\begin{figure}[tb]
\centering
\includegraphics[width=\linewidth]{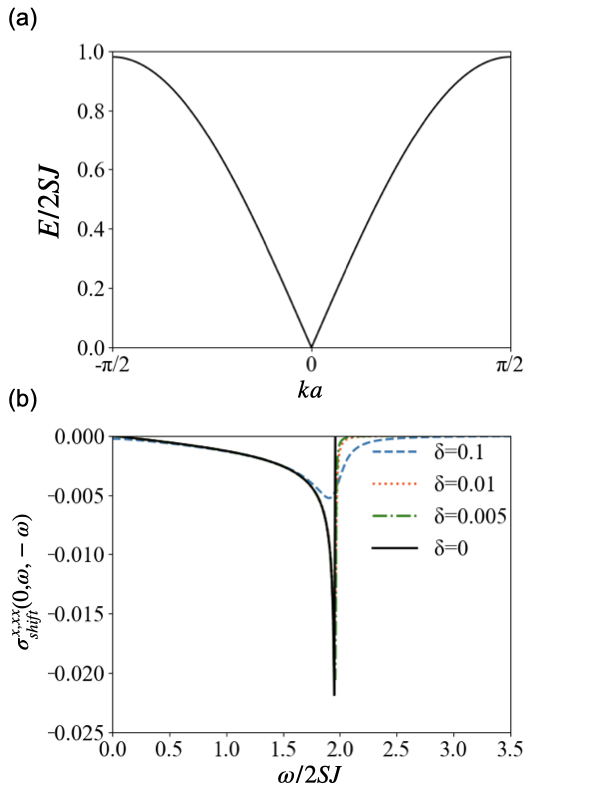}
\caption{Magnon dispersion and the magnon spin shift conductivity $\sigma^{x,xx}_{\textrm{shift}}(0,\omega,-\omega)$ for the 1D Heisenberg model with broken inversion symmetry. 
(a) Magnon dispersion. 
(b) The magnon spin shift conductivity $\sigma^{x,xx}_{\textrm{shift}}(0,\omega,-\omega)$.
The solid back curve represents the analytical result for $\delta=0$ and other curves are numerical results for $\delta=0.1,~0.01,~0.005$. 
We use parameters $2SJ=1$, $2S\delta J=0.2$, $2S\Pi=1$, and $2S\delta\Pi=0.1$.}
\label{fig:1dshift}
\end{figure}

Now, we consider the magnon spin shift conductivity $\sigma_{\textrm{shift}}^{x,xx}$. In this section, for simplicity, we consider the 1D limit, namely the case where $J_y$ and $\delta J_y$ are weak as shown in Fig.~\ref{fig:2dmodel}(b). We show the magnon dispersion and magnon spin shift conductivity in Fig.~\ref{fig:1dshift}. In the 1D limit, the excitation energy of magnons is $E_{\vb*{k}}=2S\sqrt{J^2-\delta J^2}|\sin{k_xa_x}|$ as shown in Fig.~\ref{fig:1dshift}(a), and we can derive the analytical expression as
\begin{align}
    &\sigma_{\textrm{shift}}^{x,xx}(0,\omega,-\omega)\notag\\
    &=\frac{-\delta J(J\delta\Pi-\delta J\Pi)^2\omega}{2Sa_x(J^2-\delta J^2)^2\sqrt{4(J^2-\delta J^2)-(\omega/2S)^2}}.
    \label{analyticalshift}
\end{align}
From this analytical expression, $\sigma_{\textrm{shift}}^{x,xx}(0,\omega,-\omega)$ has a peak around the resonant frequency $\omega=4S\sqrt{J^2-\delta J^2}|\sin{(k_xa_x)}|$ which is associated with two-magnon excitations around $k_x=\pi/2a_x$. Around $k_x=\pi/2a_x$, the magnon dispersion has a maximum value and the density of states of magnons is large.

We show $\sigma_{\textrm{shift}}^{x,xx}(0,\omega,-\omega)$ in Fig.~\ref{fig:1dshift}(b) with various damping $\delta$. For a large $\delta$ ($\delta=0.1J$), the peak is broadened. However, $\sigma_{\textrm{shift}}^{x,xx}(0,\omega,-\omega)$ does not show much dependence on $\delta$ in the region where $\omega$ is small ($\omega/2SJ<1.5$).
In particular, when the damping $\delta$ is smaller than $0.01$, $\sigma_{\textrm{shift}}^{x,xx}(0,\omega,-\omega)$ is almost independent of $\delta$ except for $\omega/2SJ\sim2$. Thus $\sigma_{\textrm{shift}}^{x,xx}(0,\omega,-\omega)$ is a shift current that is not depend on the damping $\delta$. 

The origin of the shift current is a broken inversion symmetry. Indeed, the analytical expression (\ref{analyticalshift}) shows that $\sigma_{\textrm{shift}}^{x,xx}=0$ when the system has a inversion symmetry, namely $\delta J=0$. In the inversion-broken system, magnons accompany nonzero polarization and excited by the light, and the shift of the magnon wave packet contribute to the spin current. 

\begin{figure*}[htb]
\centering
\includegraphics[width=\linewidth]{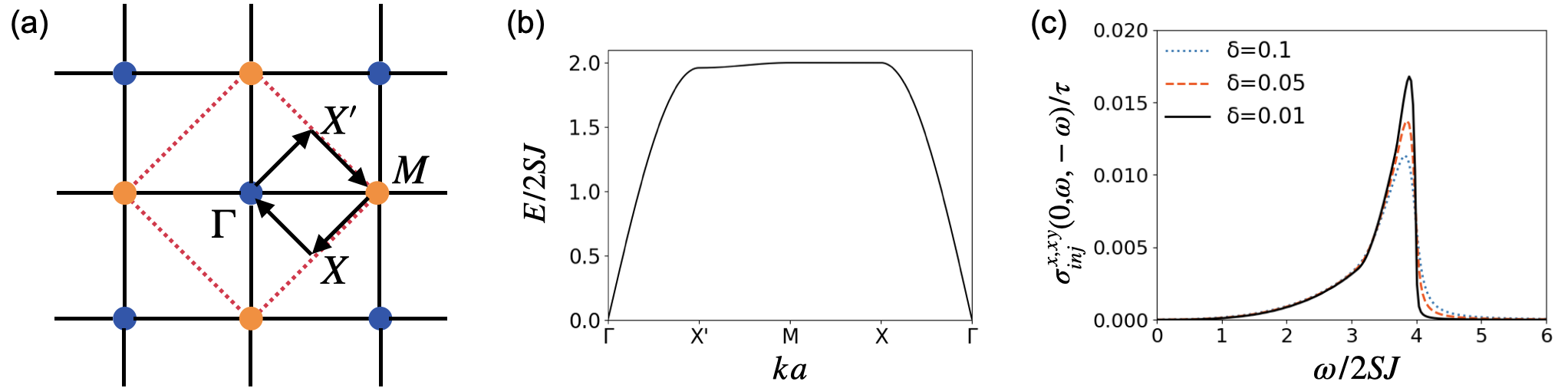}
\caption{The magnon spin injection current in the 2D Heisenberg model with broken inversion symmetry. 
(a) The reciprocal space for the 2D model. The dashed red line shows the first Brillouin zone. 
(b) Magnon dispersion along the arrow in (a). 
(c) Magnon spin injection conductivity.
We plot $\sigma^{x,xy}_{\textrm{inj}}(0,\omega,-\omega)/\tau$ for $\delta=0.1,~0.01,~0.005$ since the injection current is proportional to $\tau$. 
We use parameters $2SJ_x=2SJ_y=1$,$2S\delta J_x=2S\delta J_y=0.2$, $2S\Pi_x=1.0,2S\Pi_y=1$, $2S\delta\Pi_x=0.1$, and $2S\delta\Pi_y=0.3$.
}
\label{fig:2dinj}
\end{figure*}

\subsubsection{Magnon spin injection current}
Now, we consider the magnon spin injection current induced by the circular polarization light in 2D model depicted in Fig.~\ref{fig:2dmodel}(a). Figure~\ref{fig:2dinj} shows the magnon dispersion and the magnon spin injection current. We show the 2D model in the reciprocal space in Fig.~\ref{fig:2dinj}(a). For simplicity, we assume that $J_x=J_y$ and $\delta J_x=\delta J_y$. Thus the magnon dispersion has maximum points $E_{\vb*{k}}=4SJ_x$ at $(k_xa_x,k_ya_y)=(\pm{\pi},0),~(0,\pm{\pi})$ as shown in Fig.~\ref{fig:2dinj}(b). Around the $X^\prime$ point, a saddle point exists. Figure~\ref{fig:2dinj}(c) shows the frequency and $\delta$ dependence of the magnon spin injection conductivity divided by $\tau$ $\sigma^{x,xy}_{\textrm{inj}}(0,\omega,-\omega)/\tau$. As with the magnon spin shift current, $\sigma^{x,xy}_{\textrm{inj}}(0,\omega,-\omega)/\tau$ has a peak around the resonant frequency $\omega=4SJ$ and does not depend on $\delta$ except around the peak. However, the magnon relaxation time $\tau=1/\delta$ depends on $\delta$ and $\sigma^{x,xy}_{\textrm{inj}}(0,\omega,-\omega)$ is proportional to $\tau=1/\delta$. Thus, when the magnon relaxation time is long, $\sigma^{x,xy}_{\textrm{inj}}(0,\omega,-\omega)$ can be large and becomes the dominant contribution for the photoinduced spin current. We estimate the order of the magnon spin injection current in Sec.~\ref{sec:discussion}.

\section{Magnon spin shift current in \ce{\textit{M}_2Mo_3O_8}}
\label{sec:m2mo308}

\begin{figure*}[htb]
\centering
\includegraphics[width=\linewidth]{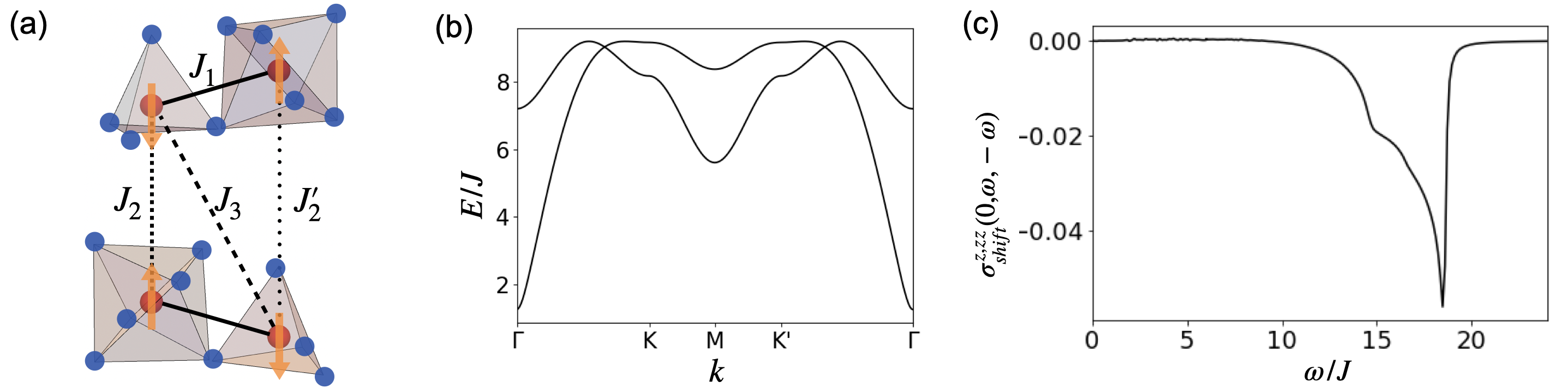}
\caption{The magnon spin shift current in the noncentrosymmetric collinear magnet \ce{Mn_2Mo_3O_8}.
(a) The magnetic structure of \ce{Mn_2Mo_3O_8}. 
Red balls represent \ce{Mn} ions and blue balls represent \ce{O} ions. The ferrimagnetic spin configuration is indicated by the orange arrows. 
(b) Magnon dispersion in the $k_z=0$ plane. A cut of the Brillouin zone in the $k_z$ plane corresponds to that for a 2D honeycomb lattice. Note that the extrema in this 1D plot does not necessarily coincide with the extrema in the full 3D dispersion.
(c) The magnon spin shift conductivity $\sigma^{z,zz}_{\textrm{shift}}(0,\omega,-\omega)$. 
We use parameters, $J_1=1$, $J_2=0.4$, $J_2^\prime=0.3$, $J_3=0.006$, $\Delta_A=-0.05$, $\Delta_B=0.008$, $\Pi_1=1$, $\Pi_2=0.2$, $\Pi_2^\prime=0.1$, $\Pi_3=0.005$, and $\delta=0.1$
}
\label{fig:M2Mo3O8}
\end{figure*}

In this section, we consider $\ce{\textit{M}_2Mo_3O_8}$ (\ce{\textit{M}}: 3d transition metal)~\cite{McCarroll1957SomeMolybdenum,McAlister1983MagneticNi,Wang2015UnveilingFe2Mo3O8,Kurumaji2017ElectromagnonFe2Mo3O8,Kurumaji2017DiagonalMn2Mo3O8} as a candidate system for the photoinduced spin current since $\ce{\textit{M}_2Mo_3O_8}$ exhibits a collinear magnetic order and an electric polarization.
The crystal structure of $\ce{\textit{M}_2Mo_3O_8}$ is composed of the alternative stacking of the $\ce{\textit{M}}$ layer and $\ce{Mo}$ layer. 
Due to the trimerization of the \ce{Mo} ions, the $\ce{Mo}$ ions are nonmagnetic, while $\ce{\textit{M}}$ ions have magnetic moments. In the magnetic \ce{\textit{M}} layer, \ce{\textit{M}} ions compose a honeycomb lattice and have two types of distinct coordination, denoted as $A$ and $B$ sites with tetrahedral and octahedral coordination of oxygen atoms, respectively. 
We show a schematic spin structure of $\ce{Mn_2Mo_3O_8}$, magnon dispersion, and the magnon spin shift conductivity in Fig.~\ref{fig:M2Mo3O8}. The ground state of $\ce{Mn_2Mo_3O_8}$ shows a ferrimagnetic spin structure~\cite{Kurumaji2017DiagonalMn2Mo3O8} as sketched in Fig.~\ref{fig:M2Mo3O8}(a). The $A$ sites and $B$ sites are inequivalent and magnitudes of spins at the $A$ sites ($S_A$) and $B$ sites ($S_B$) are generally different. However, it is reported that spontaneous magnetization asymptotically decreases to zero at zero temperature~\cite{McAlister1983MagneticNi,Kurumaji2017DiagonalMn2Mo3O8}. Thus, as far as we focus on spin current responses in low temperatures, we can approximately set $S_A=S_B$ as we do so in the following. 
Here we note that even if $S_A\neq S_B$, the formula \eqref{magnon_shift} for spin shift current in terms of the magnon shift vector is still valid since the spin current operator is expressed with the $k$ derivative of the Hamiltonian as in Eq.~(\ref{spin_current}) in the case of collinear ferrimagnets with antiferromagnetic orders.
Furthermore, $\ce{Fe_2Mo_3O_8}$ also has the ferrimagnetic phase as the ground state when a magnetic field $H~||~c$ is applied or when doped with $\ce{Zn}$~\cite{Wang2015UnveilingFe2Mo3O8,Kurumaji2017ElectromagnonFe2Mo3O8}. This ferrimagnetic order allows the polarization $P~||~c$. 
Below, we focus on the polarization along the $c$ axis and study the magnon spin shift current induced by the linear polarized light along the $c$ axis. 
We note that spin injection current under the circularly polarized light requires a nonzero polarization operator along either $a$ or $b$ direction, so that the spin shift current is the only contribution to the spin current in this case. 

Now, we consider the spin model of the $\ce{\textit{M}_2Mo_3O_8}$, following Ref.~\cite{Szaller2020MagneticMn2Mo3O8}, as
\begin{equation}
    H_s=\frac{1}{2}\sum_{i,j}J_{ij}\vb*{S}_{i}\cdot\vb*{S}_j+\Delta_A\sum_{i\in A}(S^z_i)^2+\Delta_B\sum_{i\in B}(S^z_i)^2.
    \label{M2Mo3O8_spin}
\end{equation}
Here, the bond exchange interaction $J_{ij}$ has four different nonzero values depending on the type of bonds: intralayer nearest neighbor coupling $J_1$, interlayer nearest neighbor coupling $J_2$ and $J_2^\prime$, and interlayer nearest neighbor coupling between  tetrahedrally coordinated spins $J_3$ as shown in Fig.~\ref{fig:M2Mo3O8}(a). 
While parameters of $J_2=J_2^\prime$ are adopted in  Ref.~\cite{Szaller2020MagneticMn2Mo3O8}, 
the different interspin distances associated with the tetrahedral configuration either above or below the octahedral configuration can lead to $J_2\neq J_2^\prime$ ~\cite{Streltsov2019OrderingFeZnMo3O8}, which we adopt below.

Now, we consider the polarization operator. We focus on the exchange striction mechanism and assume that exchange striction is the dominant contribution to the electric polarization.
In this case, the polarization operator can be written as
\begin{equation}
    P_s=-\frac{1}{2}\sum_{i,j}\Pi_{ij}\vb*{S}_{i}\cdot\vb*{S}_j
\end{equation}
Here, the polarization $\Pi_{ij}$ has four different nonzero values depending on the type of bonds as well as $J_{ij}$. 

According to the first-principles calculations in Ref.~\cite{Szaller2020MagneticMn2Mo3O8}, we adopt the parameters $J_1=1$, $J_2/J_1\sim0.4$, $J_2^\prime/J_1\sim0.3$, $J_3/J_1\sim0.006$, $\Delta_A/J_1\sim-0.05$, $\Delta_B/J_1\sim0.008$, and $S_A=S_B=5/2$. 
Here, for numerical stability, we use the larger value of $|\Delta_A|$ than the value of~\cite{Szaller2020MagneticMn2Mo3O8}. For the electric polarization from the exchange striction mechanism, we used the parameters $\Pi_1=1$, $\Pi_2=0.2$, $\Pi_2^\prime=0.1$, and $\Pi_3=0.005$. These parameters are chosen so that the relative magnitude for the four type of bonds are consistent with the magnitudes of the Heisenberg couplings. Here the minus sign reflects the negative polarization of \ce{\textit{M}_2Mo_3O_8} along the $c$ axis. 

Since the unit cell of the \ce{Mn_2Mo_3O_8} contains 4 \ce{Mn} ions, the magnon Hamiltonian of Eq.~(\ref{M2Mo3O8_spin}) is a $4\times4$ Hamiltonian. Thus the magnon dispersion consists of two positive energy modes and two negative energy modes. Magnon bands of the positive energy modes in the $k_z=0$ plane are shown in Fig.~\ref{fig:M2Mo3O8}(b).
Here, we note that \ce{Mn} ions compose a honeycomb lattice in $k_z=0$ plane. We show the  magnon spin shift conductivity $\sigma^{z,zz}_{\textrm{shift}}(0,\omega,-\omega)$ in Fig.~\ref{fig:M2Mo3O8}(c). The magnon spin shift conductivity $\sigma^{z,zz}_{\textrm{shift}}(0,\omega,-\omega)$ has a broad peak structure around $\omega=15\sim18J$ where the inter-band optical transition is large. As in the 1D case in Fig.~\ref{fig:1dshift}(c), the peak structure around $\omega\sim18J$ corresponds to twice the maximum value of the magnon dispersion. Furthermore, Fig.~\ref{fig:M2Mo3O8}(c) shows a shoulder structure around $\omega\sim15J$ which corresponds to twice the minimum values of the top magnon band.

We note on the magnitude of $\delta$. Because of computational complexity in 3D systems, we used a relatively large value of $\delta=0.1$. As we show in Fig.~\ref{fig:1dshift}(c),  $\sigma_{\textrm{shift}}(0,\omega,-\omega)$ is expected to show a qualitatively same behavior with those for smaller $\delta$, except around the peak. The peak structure is expected to be sharper for smaller $\delta$ and $\sigma^{z,zz}_{\textrm{shift}}(0,\omega,-\omega)$ in the peak region becomes larger by reducing $\delta$.

\section{Discussion}
\label{sec:discussion}
We have derived the expression of the second order magnon spin conductivity and clarify the relationship between the magnon spin shift current and the shift vector, for collinear magnets. In noncentrosymmetric systems, the electric field can generally excite electromagnons, where photoinduced magnons exhibit a positional shift and induce spin polarization. Furthermore, we have studied the magnon spin injection current which is proportional to the relaxation time $\tau$. 
Our numerical calculation demonstrates that the collinear spin systems with an electric polarization from the exchange striction mechanism support the magnon spin shift current and injection current.

The magnon spin current can be experimentally observed using setups with Kerr rotation or Faraday effect, or a two-terminal setup with a non-centrosymmetric magnetic insulator sandwiched between two metallic leads, as suggested in the previous theoretical proposasl~\cite{Ishizuka2019TheoryInsulators}. 
The nonlinear magnon spin conductivity and the strength of the electric field to support a realistic value of spin current, which is of the order of $J_s=10^{-16}$~J/cm$^2$~\cite{Ishizuka2019RectificationWaves,Ishizuka2019TheoryInsulators,Hirobe2017One-dimensionalCurrents},
can be estimated as follows.

First, we consider the magnitude of the magnon spin shift conductivity $\sigma_{\textrm{shift}}$. One candidate material for magnon spin current is $\ce{\textit{M}_2Mo_3O_8}$, as we detailed in Sec.~\ref{sec:m2mo308}. 
The parameters for $\ce{Mn_2Mo_3O_8}$ are given as follows: The lattice constants in the in-plane and perpendicular directions are $a_a\sim 6$~\AA  ~and $a_c\sim 10$~\AA, respectively~\cite{McCarroll1957SomeMolybdenum}. The exchange interaction $J_1$ is $0.8$~meV~\cite{Szaller2020MagneticMn2Mo3O8}. The spin-induced spontaneous polarization of $\ce{\textit{M}_2Mo_3O_8}$ at low temperature $P-P_{T_c}\sim-1500$~{\textmu}C/m$^2$. Assuming that $\Pi_1$ gives a dominant term for the electric polarization operator, and given that the number of intralayer nearest neighbor bonds is 6 in the unit cell, we estimate $6\Pi_1/V\sim-1500$~{\textmu}C/m$^2$ from the above value of $P-P_{T_c}$, where $V$ is a volume of the unit cell. The value of $\Pi_1/V\sim-250$~{\textmu}C/m$^2$ is consistent with that of $\Pi_1$ in \ce{Fe_2Mo_3O_8} calculated from the first-principles calculations~\cite{Solovyev2019MicroscopicFe2Mo3O8}. 
From these values and our numerical result in Fig.~\ref{fig:M2Mo3O8}(c), 
we obtain $\sigma_{\textrm{shift}}\sim5\times10^{-24}$~C$^2$/J,
which indicates that applying an ac electric field of $E\sim10^4$~V/cm leads to an experimentally detectable magnon spin current of $J_s\sim5\times10^{-16}$~J/cm$^2$.

Next, let us estimate the order of the magnon spin injection conductivity $\sigma_{\textrm{inj}}^{\mu,\alpha\beta}(0,\omega,-\omega)$. The magnon spin injection current can be greater than the shift current when the relaxation time is long. 
We use the antiferromagnetic spin model in Sec.~\ref{sec:2d} for the estimation. 
We assume the energy gap $\varepsilon_{ab}\simeq J_x=J_y$ of the order of $\simeq 10$~meV, the exchange striction with $\Pi_x/V=\Pi_y/V=250$~{\textmu}C/m$^2$, 
and the lattice constant $a\sim5$\AA. 
The relaxation time $\tau$ can be estimated from $\tau=1/\alpha\omega$, where $\alpha$ is Gilbert damping constant. 
Using the parameter $\alpha=2\times 10^{-4}$ for the antiferromagnet \ce{NiO}~\cite{Kampfrath2011CoherentWaves} and the resonant frequency $\omega\sim\varepsilon_{ab}\simeq 10$~meV, 
we estimate $\tau$ as $3\times 10^{-10}$~s. 
From these values and the result in Fig.~\ref{fig:2dinj}(c), we obtain $\sigma_{\textrm{inj}}\sim3\times10^{-22}$~C$^2$/J. If we apply the ac electric field $E\sim10^4$~V/cm, the estimated magnon spin current amounts to $J_s\sim3\times10^{-14}$~J/cm$^2$, which shows that the magnon spin injection current is much larger than the magnon spin shift current. 
Since the energy scale of the magnon is around $10$~meV which corresponds to a few THz in the frequency range,
the magnon spin current is well feasible for experimental detection with an irradiation of the THz light field of the order of  $10^3\sim10^4$~V/cm.

Finally, we comment on the validity of our expression of the magnon spin shift current in terms of the shift vector.
In deriving Eq.~(\ref{magnon_shift}), we relied on the fact that the spin current can be written as the $k$-derivative of the Hamiltonian, which is true for collinear magnets. However, in the noncollinear magnets, the magnon Hamiltonian contains three-magnon terms and the spin current operator cannot be written as the $k$-derivative of the Hamiltonian. Furthermore, in the noncollinear magnets, there are contributions of the single-magnon resonance to the spin conductivity, which is not considered in our research.
Thus our geometric description of the magnon shift current is only valid for collinear magnets, and its extension for more general cases is left for a future work.

\begin{acknowledgments}
We thank Hosho Katsura for useful discussions.
This work was supported by 
JSPS KAKENHI Grant 20K14407 (S.K.), 
JST CREST (Grant No. JPMJCR19T3) (S.K., T.M.),
and JST PRESTO (Grant No. JPMJPR19L9) (T.M.).
K.F. was supported by the Forefront Physics and Mathematics program to drive transformation (FoPM).
\end{acknowledgments}

\appendix

\section{Derivation of the nonlinear spin conductivity in Eq.~(\ref{2ndsigma})}
\label{Appendix}
In this appendix, we present a derivation of the nonlinear spin conductivity in Eq.~(\ref{2ndsigma}) from a standard perturbation theory. To this end, we introduce the Matsubara Green function of the magnon  
\begin{align}
G_{a,b}(\tau,\vb*{k})&=-\langle\mathcal{T}[(\Psi_{\vb*{k}}(\tau))_a(\Psi^\dagger_{\vb*{k}}(0))_b]\rangle\\
&=-\sum_{c,d}(V_{\vb*{k}})_{ac}(V^\dagger_{\vb*{k}})_{db}\langle\mathcal{T}[(\Phi_{\vb*{k}}(\tau))_c(\Phi^\dagger_{\vb*{k}}(0))_d]\rangle
\end{align}
Here, $\tau$ is the imaginary time and $\mathcal{T}$ is the time-ordered product. Since the transformed operator $\Phi_{\vb*{k}}$ is the basis of the diagonalized Hamiltonian, we obtain
\begin{align}
    &-\frac{\partial}{\partial\tau}\langle\mathcal{T}[(\Phi_{\vb*{k}}(\tau))_c(\Phi^\dagger_{\vb*{k}}(0))_d]\rangle\notag\\
    &=B_{c,d}-B_{cc}(E_{\vb*{k}})_{cc}\langle\mathcal{T}[(\Phi_{\vb*{k}}(\tau))_c(\Phi^\dagger_{\vb*{k}}(0))_d]\rangle.
\end{align}
By using the Fourier transformation, the Matsubara Green function can be written as
\begin{align}
G_{a,b}(i\omega,\vb*{k})
&=-\sum_{c,d}(V_{\vb*{k}})_{ac}(V^\dagger_{\vb*{k}})_{db}\frac{B_{cd}}{i\omega-(BE_{\vb*{k}})_{cd}}\notag\\
&=\left(V_{\vb*{k}}\frac{B}{i\omega-BE_{\vb*{k}}}V^\dagger_{\vb*{k}}\right)_{ab}.
\end{align}
Thus we obtain
\begin{align}
G(i\omega,\vb*{k})&=V_{\vb*{k}}(i\omega -\varepsilon_{\vb*{k}})^{-1}BV_{\vb*{k}}^{\dagger}\notag\\
&=V_{\vb*{k}}(i\omega -\varepsilon_{\vb*{k}})^{-1}V_{\vb*{k}}^{-1}B\\
&=(i\omega-BH_{\vb*{k}})^{-1}B,\label{greenfunc}
\end{align}
where we used the diagonalized representation as in Eq.~(\ref{varepsilon}). 

Now we express the second order conductivity by using the Green function.
First, we consider the $\vb*{J}_{1}$ term of the spin current. The contribution of the $\vb*{J}_{1}$ term to the second order magnon spin conductivity $\sigma^{\mu,\alpha\beta}(i\Omega_m+i\Omega_n,i\Omega_m,i\Omega_n)$ is given by 
\begin{widetext}
\begin{align}
\sigma^{\mu,\alpha\beta}_1(i\Omega_m+i\Omega_n,i\Omega_m,i\Omega_n)
=&\int \frac{d\vb*{k}}{(2\pi)^d}\int_c\frac{dz}{2\pi i} F(z)\textrm{Tr}\left[
J^\mu_{1\vb*{k}}G(z+i\Omega_m+i\Omega_n,\vb*{k})\Pi^\alpha_{\vb*{k}}G(z+i\Omega_m,\vb*{k})\Pi^\beta_{\vb*{k}}G(z,\vb*{k})
\right]\notag\\
&+(\alpha,\Omega_m\leftrightarrow\beta,\Omega_n)
,
\end{align}
where $\sigma^{\mu,\alpha\beta}_1(i\Omega_m+i\Omega_n,i\Omega_m,i\Omega_n)$ is the contribution of $\vb*{J}_{1}$ to $\sigma^{\mu,\alpha\beta}(i\Omega_m+i\Omega_n,i\Omega_m,i\Omega_n)$. Here, $\Omega_m$ is a Matsubara frequency of boson and $F(z)=\frac{\beta}{2}\coth{\frac{\beta z}{2}}$ is a Matsubara weighting function. We rewrite the Green funtion by using Eq. (\ref{greenfunc}) and perform the Matsubara frequency summation, we obtain
\begin{align}
\sigma^{\mu,\alpha\beta}_1(i\Omega_m+i\Omega_n,i\Omega_m,i\Omega_n)
=&\int \frac{d\vb*{k}}{(2\pi)^d}\int_c\frac{dz}{2\pi i} F(z)\textrm{Tr}\left[
J^\mu_{1\vb*{k}}(z+i\Omega_m+i\Omega_n -BH_{\vb*{k}})^{-1}B\Pi^\alpha_{\vb*{k}}(z+i\Omega_m -BH_{\vb*{k}})^{-1}\right.\notag\\
&\left.\times B\Pi^\beta_{\vb*{k}}(z -BH_{\vb*{k}})^{-1}B
\right]+(\alpha,\Omega_m\leftrightarrow\beta,\Omega_n)\notag\\
=&\int \frac{d\vb*{k}}{(2\pi)^d}\sum_{a,b,c}\int_c\frac{dz}{2\pi i} F(z)
\frac{\tilde{J}^\mu_{1ac}\tilde{\Pi}^\alpha_{ba}\tilde{\Pi}^\beta_{cb}}{(z+i\Omega_m+i\Omega_n-\varepsilon_c)(z+i\Omega_m-\varepsilon_b)(z-\varepsilon_a)}\notag\\
&+(\alpha,\Omega_m\leftrightarrow\beta,\Omega_n)\notag\\
=&-\int\frac{d\vb*{k}}{(2\pi)^d}\sum_{a,b,c}\frac{\tilde{J}^\mu_{1ac}\tilde{\Pi}^\alpha_{ba}\tilde{\Pi}^\beta_{cb}}{i\Omega_m+i\Omega_n+\varepsilon_{ac}}\left(\frac{f_{ab}}{\varepsilon_{ab}+i\Omega_m}+\frac{f_{cb}}{\varepsilon_{bc}+i\Omega_n}\right)+(\alpha,\Omega_m\leftrightarrow\beta,\Omega_n).
\end{align}
Next, we consider the contribution of the $\vb*{J}_{2}$ term which is a first-order term with respect to the electric field. As in the case of $\vb*{J}_1$ term, we can write $\sigma^{\mu,\alpha\beta}_2(i\Omega_m+i\Omega_n,i\Omega_m,i\Omega_n)$ as the contribution of $\vb*{J}_{2}$ to $\sigma^{\mu,\alpha\beta}(i\Omega_m+i\Omega_n,i\Omega_m,i\Omega_n)$ and obtain
\begin{align}
\sigma^{\mu,\alpha\beta}_2(i\Omega_m+i\Omega_n,i\Omega_m,i\Omega_n)
=&\int \frac{d\vb*{k}}{(2\pi)^d}\int_c\frac{dz}{2\pi i} F(z)\textrm{Tr}\left[
J^{\mu,\alpha}_{2\vb*{k}}G(z,\vb*{k})\Pi^\beta_{\vb*{k}}G(z+i\Omega_m,\vb*{k})\right]\notag\\
&+(\alpha,\Omega_m\leftrightarrow\beta,\Omega_n)\notag\\
=&\int \frac{d\vb*{k}}{(2\pi)^d}\int_c\frac{dz}{2\pi i} \sum_{ab}F(z)
\frac{\tilde{J}^{\mu,\alpha}_{2ab}\tilde{\Pi}^\beta_{ba}}{(z+i\Omega_m-\varepsilon_b)(z-\varepsilon_a)}+(\alpha,\Omega_m\leftrightarrow\beta,\Omega_n)\notag\\
=&-\int\frac{d\vb*{k}}{(2\pi)^d}\sum_{ab}\tilde{J}^{\mu,\alpha}_{2ab}\tilde{\Pi}^\beta_{ba}\frac{f_{ab}}{i\Omega_m+\varepsilon_{ab}}+(\alpha,\Omega_m\leftrightarrow\beta,\Omega_n).
\end{align}
The second order magnon spin conductivity can be obtained by combining these two contributions as
\begin{align}
  \sigma^{\mu,\alpha\beta}(i\Omega_m+i\Omega_n,i\Omega_m,i\Omega_n)=\sigma^{\mu,\alpha\beta}_1(i\Omega_m+i\Omega_n,i\Omega_m,i\Omega_n)+\sigma^{\mu,\alpha\beta}_2(i\Omega_m+i\Omega_n,i\Omega_m,i\Omega_n),
\end{align}
which directly leads to Eq.~$(\ref{2ndsigma})$ by performing analytic continuation of the Matsubara frequency $\Omega_m\rightarrow\omega+i\delta$ and $\Omega_n\rightarrow-\omega+i\delta$.
\end{widetext}

\nocite{*}

\bibliography{references}

\end{document}